\documentclass[journal]{IEEEtran}

\usepackage[T1]{fontenc}
\usepackage[utf8]{inputenc}

\usepackage{amsthm}

\usepackage{bm}

\usepackage{import}
\usepackage{currfile}

\usepackage{tudacolors}

\usepackage[numbers,sort&compress]{natbib}

\usepackage[scaled=0.94]{helvet}     \usepackage{titlesec}   \usepackage{multicol}   \usepackage{booktabs}   \usepackage{float}      \usepackage{framed}     \usepackage{eurosym}    \usepackage{siunitx}    \usepackage{enumitem}

\usepackage[]{subfig}

\usepackage{xcolor}
\usepackage{graphicx}
\usepackage{pgfplots}

\usepackage{amsmath}
\usepackage{amssymb}
\usepackage{mathtools}
\usepackage{bm}

\usepackage[bookmarks=true,
            linkcolor=darkblue,
            citecolor=darkblue,
            urlcolor=darkblue,
            colorlinks=true,
            breaklinks=true]{hyperref}

\usepackage[labelfont=bf,
            font=small]{caption}

\usepackage[most]{tcolorbox} 

\usepackage{tudacolors}

\usepackage{glossaries}
\usepackage[capitalise]{cleveref}
 \DeclareMathAlphabet\mathbfcal{OMS}{cmsy}{b}{n}
\definecolor{darkblue}{rgb}{0,0,.7}

\newtheorem{definition}{Definition}

\newtheorem{theorem}{Theorem}
\newtheorem{corollary}{Corollary}

\colorlet{myblue}{TUDa-1b}
\colorlet{myred}{TUDa-9b}
\colorlet{mygreen}{TUDa-6b}
\colorlet{myviolet}{TUDa-11b}
 \DeclareMathOperator{\E}{\mathsf{E}}
\DeclareMathOperator{\Var}{\mathsf{Var}}

\DeclareMathOperator*{\argmax}{\mathrm{arg\,max}}
\DeclareMathOperator{\softmin}{\mathrm{softmin}}

\newcommand{\obs}{\mathbf{x}}
\newcommand{\obsRV}{\mathbf{X}}
\newcommand{\Hyp}[1][i]{\mathrm{H}_{#1}}
\newcommand{\param}[1][i]{\theta_{#1}}
\newcommand{\paramRV}[1][i]{\Theta_{#1}}
\newcommand{\discrParam}[2][i]{\param[#1]^{[#2]}}

\newcommand{\policy}{\pi}

\newcommand{\dec}{\delta}
\newcommand{\est}[1][i]{\hat\theta_{#1}}

\newcommand{\policyOpt}{\pi^\star}
\newcommand{\decOpt}{\delta^\star}
\newcommand{\estOpt}[1][i]{\hat\theta_{#1}^\star}

\newcommand{\detErr}[1][i]{\alpha^{#1}}
\newcommand{\estErr}[1][i]{\beta^{#1}}

\newcommand{\detErrMax}[1][i]{\bar{\alpha}^{#1}}

\newcommand{\detCost}[1][i]{\lambda_{#1}}
\newcommand{\estCost}[1][i]{\mu_{#1}}

\newcommand{\dInt}{\mathrm{d}}
\newcommand{\given}{\,|\,}
\newcommand{\Given}{\,\Big|\,}

\newcommand{\iid}{\overset{\text{iid}}{\sim}}

\newcommand{\stateSpace}{E}
\newcommand{\stateSpaceParam}[1][i]{\stateSpace_{\paramRV[#1]}}
\newcommand{\stateSpaceHyp}[1][i]{\stateSpace_{\Hyp[#1]}}
\newcommand{\stateSpaceObs}{\stateSpace_{\obsRV}}

\newcommand{\NPcost}{J^\text{NP}}
\newcommand{\NPcostUnconstr}{{J}^\text{NP}_\text{u}}

\newcommand{\NPcostUnconstrDiscr}{\tilde{J}^\text{NP}_\text{u}}
\newcommand{\BayesCost}{J^\text{B}}

\newcommand{\arbConst}{c}

\newcommand{\densSet}{\mathbfcal{P}}

\newcommand{\detDens}[1][i]{p^\text{D}_{#1}}
\newcommand{\estDens}[1][i]{p^\text{E}_{#1}}

\newcommand{\detDensSetLFD}[1][i]{\bm{q}_{{#1}}^\text{D}}

\newcommand{\allDensLFD}{\bm{q}}
\newcommand{\allDens}{\bm{p}}

\newlength{\imgWidthSingle}
\newlength{\imgWidthDouble}
 \newacronym{ao}{AO}{asymptotically optimal}
\newacronym{lfd}{LFD}{least favorable distribution}
\newacronym{KL}{KL}{Kullback-Leibler}
\newacronym{NP}{NP}{Neyman-Pearson}
\newacronym{mse}{MSE}{mean-squared error}
\newacronym{pdf}{pdf}{probability density function}
\newacronym{mmse}{MMSE}{minimum mean-squared error}
\newacronym{jde}{JDE}{joint detection and estimation}
 \definecolor{my_red}{cmyk/RGB/HTML}{0,1,.9,0/230,0,26/E6001A}

\definecolor{my_blue}{cmyk/RGB/HTML}{1,.6,0,0/0,90,169/005AA9}

\definecolor{my_green}{cmyk/RGB/HTML}{1,0,0.6,0/0,157,129/009D81}

\definecolor{my_violet}{cmyk/RGB/HTML}{0.4,1,0,0/166,0,132/A60084}

\pgfplotsset{areaLegend/.style={
  legend image code/.code={\fill[#1] (0cm,-0.1cm) rectangle (0.6cm,0.1cm);}
  }
}

\pgfplotsset{compat=1.15}
 
\author{Dominik Reinhard,~\IEEEmembership{Senior Member,~IEEE,}
        Michael Fau\ss{},~\IEEEmembership{Member,~IEEE,}
        and~Abdelhak~M.~Zoubir,~\IEEEmembership{Life~Fellow,~IEEE}\thanks{Dominik Reinhard and Abdelhak M. Zoubir are with the Signal Processing Group, Technische Universit\"at Darmstadt, Merckstra\ss{}e 25, 64283 Darmstadt, Germany. E-mail: \{reinhard, zoubir\}@spg.tu-darmstadt.de}
\thanks{Michael Fau\ss{} is with with the Department of Electrical Engineering, Princeton University, Princeton, NJ 08544, USA. E-mail: mfauss@princeton.edu}
}

\title{Minimax Optimal Procedures for Joint Detection and Estimation}

\begin{document}
  
 \maketitle
    \begin{abstract}
We investigate the problem of jointly testing a pair of composite hypotheses and, depending on the test result, estimating a random parameter under distributional uncertainties.
Specifically, it is assumed that the distribution of the data given the parameter of interest, is subject to uncertainty.
Both, a Bayesian formulation and a Neyman-Pearson-like formulation, are considered.
It is shown that the optimal policy induces an $f$-similarity that must be maximized to identify the least favorable distributions.
Besides the general results, the implementation is investigated using a band-type uncertainty model.
For designing the minimax procedures, existing algorithms are modified to increase convergence speed while maintaining numerical stability.
The proposed theory is supplemented by numerical results for both formulations.
\end{abstract}

     \section{Introduction}
Detection and estimation are fundamental signal processing tasks that occur in a wide range of applications.
However, in many applications both inference tasks are intrinsically coupled.
That is, the outcome of the hypothesis test would influence the estimation and vice versa.
In radar, for example, one wishes to detect a target and if a target is declared to be present, one aims at estimating some parameters, such as position or velocity of the target \cite{tajer2010optimal,chen2018impact}.
In this example, the outcome of the hypothesis test as well as the outcome of the estimation task are of primary interest and should come with statistical performance guarantees.
Although one could combine an optimal detector and an optimal estimator to approach this problem, the result might not be overall optimal and, therefore, the problem of coupled detection and estimation has to addressed jointly \cite{moustakides2012joint}, which is referred to as \gls{jde} in the sequel.
Besides the scenario mentioned above, coupled detection and estimation occur in various situations.
In communication, for example, one has to detect the presence of a signal and estimate the channel once a signal is detected \cite{jan2018iterative}.
Detecting the primary user by the secondary user and estimating possible inference is a main task in cognitive radio \cite{yilmaz2014sequential}.
Additionally, the problem of \gls{jde} arises in biomedical engineering \cite{chaari2012fast, makni2008fully}, visual inference \cite{vo2010joint}, changepoint detection \cite{boutoille2010hybrid} or speech detection \cite{momeni2015joint}.
The problem of \gls{jde} has first been studied in the 1960s \cite{Middleton1968} and was later extended to the $M$-ary scenario \cite{Fredriksen1972}.
Subsequently, interest in this topic subsided, but regained more attention in the new millennium \cite{moustakides2012joint,yilmaz2014sequential,yilmaz2015sequential,yilmaz2016sequential,jan2018iterative,chaari2012fast,makni2008fully,momeni2015joint,vo2010joint,boutoille2010hybrid,Lan2025,Zhu2025}.
Especially when solving the problem of \gls{jde} in a sequential framework, one is able to control the error probabilities and the estimation error levels individually \cite{reinhard2018,reinhard2020,reinhard2022asymptotically}, which is highly interesting for applications in which a flexible performance control is required.

In statistical inference, often a precise statistical model is required that matches the true data generating procedure.
However, most commonly used models do not perfectly reflect reality \cite{box1979robustness}.
The assumption of Gaussian distributed noise, for example, can be justified theoretically, but does not always hold in practice \cite{zoubir2012robust}.
Small deviations from the assumed model may heavily degrade the performance.
In the 1960s, robust statistics arose to overcome this problem.
In robust statistics, one aims to develop methods whose performance is close to optimal under the nominal model while tolerating a certain, predefined amount of distributional uncertainty without breaking down.
The first formal treatment of what we call statistical robustness today dates back to the seminal work of Huber on robust estimation \cite{huber1964robust_estimation} and robust detection \cite{huber1965robust_detection}.
Since then, robust methods for detection and estimation as well as other signal processing tasks have been studied extensively.
Recent overviews on robust detection and estimation are given in \cite{fauss2021} and \cite{zoubir2012robust,zoubir2018robust}, respectively.

However, the research on distributional uncertainties or robustness in the context of \gls{jde} is scare.
In \cite{reinhard2016approach}, we investigated the problem of sequential \gls{jde} and combined classical results from robust binary hypothesis with a variance maximizing density to introduce robustness.
By applying robust estimators, we have developed robust methods for sequential \gls{jde} in distributed sensor networks \cite{reinhard2021distributed}.
A restricted Bayes approach to tackle uncertainty about the prior probabilities of the hypotheses has been proposed in \cite{dulek2018restricted,bayram2016joint}.

In this work, we consider the problem of \gls{jde}, where the densities conditioned on the parameter of interest are subject to uncertainty.
More precisely, this problem is investigated under a Bayesian and a \gls{NP}-like formulation.
Here, the aim is to find a set of error maximizing distributions, i.e., the \glspl{lfd}, as well as a corresponding policy.

\subsection{Contributions and Overview}
The main contributions of this work can be summarized as follows.
\begin{enumerate}
    \item \textbf{Problem formulations:} We investigate minimax procedures under the following objectives.
    \begin{enumerate}
        \item Bayesian formulation: minimize weighted sum of detection and estimation errors.
        \item \gls{NP}-like formulation: minimize  weighted sum of estimation errors and constraining the error probabilities.
    \end{enumerate}
    \item \textbf{Solutions} under both objectives:
    \begin{enumerate}
    \item Bayesian formulation: we show that the \glspl{lfd} are characterized as the maximizers of an appropriate $f$-similarity.
    \item NP-like formulation: we derive a set of sufficient conditions that guarantee minimax optimality of the resulting procedure. In particular:
        (i) the \glspl{lfd} are identified as the maximizers of an $f$-similarity induced by the optimal policy and (ii) we provide conditions on the optimal policy that ensure the constraints on the error probabilities are satisfied.
    \end{enumerate}
    \item \textbf{Connection to existing theory:} We discuss how the obtained results relate to and are consistent with existing results in robust statistics.
    \item \textbf{Practical implementation:} We adapt existing algorithms so that minimax \gls{jde} procedures can be efficiently and numerically stably designed when band-type uncertainty classes are used.
\item \textbf{Numerical validation:} The theoretical findings are illustrated by a numerical example for both the Bayesian and the \gls{NP}-like formulations.
\end{enumerate}

\subsection{Notations}
Lower and upper case letters denote respectively deterministic and random quantities.
Normal font symbols, bold symbols and calligraphic symbols represent scalars, vectors and sets, respectively.
For a vector $\bm a$, $\bm a^\top$ denotes its transpose and $[\bm a]_i$ its $i$th element.
The expected value and the variance of a random variable $X$ are denoted by $\E_p[X]$ and $\Var_p[X]$, respectively, where the subscript denotes the density used for calculating the moments.
For the sake of a compact notation, the integration domain is not stated explicitly if the integral is taken over the whole domain, e.g., $\E[X] = \int_{-\infty}^\infty x\cdot p(x)\dInt x = \int x\cdot p(x)\dInt x$.

\subsection{Outline}
The work is structured as follows.
In \cref{sec:minimax}, we review briefly some fundamentals of minimax procedures.
Subsequently, the system model and a detailed problem statement are presented in \cref{sec:problem}.
The characterization of the solution under the Bayesian and the \gls{NP}-like formulation is provided in \cref{sec:Bayes} and \cref{sec:NP}, respectively.
A thorough discussion of the theoretical results can be found in \cref{sec:discussion}.
An algorithm for computing the densities is given in \cref{sec:calc_distr}, followed by numerical results for the Bayesian and the \gls{NP}-like formulation in \cref{sec:num_results}.
Conclusions are drawn in \cref{sec:conclusion}.     \section{Minimax Statistical Inference}\label{sec:minimax}
This section provides the fundamentals of minimax statistical inference used in this work.
Designing minimax optimal procedures can be boiled down to finding a solution of
\begin{align}\label{eq:minimax}
    \sup_{u\in \mathcal{U}} \min_{v\in \mathcal{V}} J(u,v) = \min_{v\in \mathcal{V}} \sup_{u\in \mathcal{U}}  J(u,v)\,,
\end{align}
where $J(u,v)$ is some cost function.
In the equation above, $u$ represents the statistical procedure, e.g., it can be a decision rule in the context of hypothesis testing or an estimator in the context of parameter estimation.
The variable $v$ on the other hand represents the quantity which is subject to uncertainty, which could, for example, be the distribution of the data itself or just a single parameter of the distribution.
For \cref{eq:minimax} to hold with equality, several conditions on the cost function as well as on the spaces $\mathcal{U}$ and $\mathcal{V}$ have to be fulfilled.
If equality holds, a saddle point of the form
\begin{align*}
    J(u,v^\star) \leq J(u^\star,v^\star) \leq J(u^\star,v), \quad \forall u\in \mathcal{U}, v\in \mathcal{V}
\end{align*}
exists, where $u^\star$ and $v^\star$ are the solutions of \cref{eq:minimax}.
Hence, the validity of \cref{eq:minimax} is a consequence of the existence of a saddle point.

The existence of minimax optimal procedures is stated in Sion's famous minimax theorem.
\begin{theorem}[Sion's minimax theorem \cite{Sion1958}]\label{theo:minimax}
Let $\mathcal{U}$ and $\mathcal{V}$ be convex sets.
 Then, \cref{eq:minimax} holds
if
\begin{enumerate}
 \item $J(\cdot, v)$ is upper semi-continuous and convex in $u$ for every fixed $v\in \mathcal{V}$
 \item $J(u, \cdot)$ is lower semi-continuous and concave in $v$ for every fixed $u\in \mathcal{U}$
 \item either $\mathcal{U}$ or $\mathcal{V}$ is compact.
\end{enumerate}
\end{theorem}

\begin{figure}[tp]
    \centering
    \includegraphics{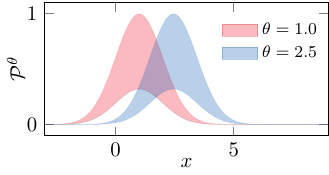}
    \caption{Illustrative example of a density band model. The set of feasible distributions $\mathcal{P}^\theta$ depends on the location parameter $\theta$.}
    \label{fig:band_model_example}
\end{figure}

Quantifying the model uncertainties is an essential step prior to the design of robust methods.
There exist various ways to quantify distributional uncertainties in the literature.
In robust statistics, one defines the set of feasible distributions as a neighborhood around some nominal model.
This can, for example, be done by placing an $f$-divergence ball around a nominal model, such as a \gls{KL} divergence ball \cite{gul2017minimax} or an $\alpha$-divergence ball \cite{gul2016robust}.
Additionally, the neighborhood may be defined as a mixture of the nominal and an arbitrary contaminating distribution, which is referred to as the $\varepsilon$-contamination model \cite{huber1965robust_detection,huber1964robust_estimation}.
A different approach, introduced by Kassam \cite{Kassam1981}, is to model the set of feasible distributions using the band model.
Here, all densities are assumed to lie between a pre-defined upper and lower bound, see \cref{fig:band_model_example} for illustration.
Though this band model can, in principle, be placed around a nominal distribution, it is in general non-parametric.
If the upper bound tends to infinity, this model coincides with the $\varepsilon$-contamination model \cite{fauss2016old}.
Furthermore, the band model can be used to capture local uncertainties that vary over the state space \cite{pambudi2020minimax}.

Statistical similarity measures, such as $f$-divergences \cite{csiszar1963informationstheoretische,morimoto1963markov,ali1966general,cha2007comprehensive} and $f$-dissimilarities \cite{GyorfiNemetz1975,gyorfi1978f,GyorfiNemetz1977} as their natural extension to more than two distributions play a central role in statistics and in particular in robust hypothesis testing.
More precisely, the \glspl{lfd} are given as the minimizer of \emph{all} $f$-divergences in binary hypothesis testing \cite{huber1973minimax} and as the minimizer of a particular $f$-dissimilarity in more complex setups such as, $M$-ary testing \cite{fauss2021} or sequential hypothesis testing \cite{fauss2016,fauss2020minimax,fauss2021}.

Next, a formal definition of $f$-(dis)similarities is given.
\begin{definition}[$f$-(dis)similarity \cite{gyorfi1978f}]\label{def:f-sim}
Let $f(s_1,\ldots,s_N)$ be a continuous, concave (convex), homogeneous\footnote{We use \cite[Definition 2]{gyorfi1978f} as the definition for homogeneity throughout this paper.} function defined on $\mathbb{R}_{\geq0}$ and let $P_1,\ldots,P_N$ be probability distributions with densities $p_1(\obs),\ldots, p_N(\obs)$.
Then, the function
\begin{align*}
 D_f(P_1,\ldots,P_N) = \int f(p_1(\obs),\ldots, p_N(\obs))\dInt \obs
\end{align*}
is called $N$-dimensional $f$-similarity ($f$-dissimilarity).
\end{definition}     \section{Problem Formulation}\label{sec:problem}
In what follows, the system model used in this work is presented.
Subsequently, some fundamentals of \gls{jde}, such as performance measures, are introduced, followed by problem formulations for optimal joint detection and estimation and the minimax optimal problem formulation.
\subsection{System Model}
Let $\obsRV$ be a random variable that can be generated under two different hypotheses $\Hyp[0]$ and $\Hyp[1]$.
The occurrence of the hypotheses is random with known prior probabilities $P(\Hyp[0])$ and $P(\Hyp[1])$, respectively.
Under each hypothesis, the distribution of the data may depend on a random parameter $\paramRV[i]$, $i\in\{0,1\}$, with known distribution.
The triplet ($\obsRV$, $\paramRV$, $\Hyp$) is defined on the parameter space $\stateSpaceObs\times\stateSpaceParam\times\stateSpaceHyp$.

Hence, the model can be written as
\begin{align}\tag{M1}\label{eq:model}
    \begin{split}
        \Hyp[0]: & \quad  \obsRV\given\param[0] \iid P(\obs\given\Hyp[0],\param[0])\,, \quad \paramRV[0] \sim P(\param[0]\given\Hyp[0])\,,\\
        \Hyp[1]: & \quad  \obsRV\given\param[1] \iid P(\obs\given\Hyp[1],\param[1])\,, \quad \paramRV[1] \sim P(\param[1]\given\Hyp[1])\,.
    \end{split}
\end{align}

For the model \cref{eq:model}, there are three statistical quantities which could be subject to uncertainty, namely, the conditional distribution of the data $P(\obs\given\Hyp,\param)$, $i\in\{0,1\}$, the prior distribution of the parameters $P(\param\given\Hyp)$, $i\in\{0,1\}$, and the prior probabilities of the hypotheses $P(\Hyp)$, $i\in\{0,1\}$.
In this work, we focus on uncertainty in the conditional distributions of the data, i.e., the likelihoods.
More precisely, we assume that for every $\param\in\stateSpaceParam$, the likelihood belongs to some convex uncertainty set $\mathcal{P}_{i,\param}$.
Mathematically, that is
\begin{align*}
 P(\obs\given\Hyp[i],\param[i]) \in \mathcal{P}_{i,\param}\quad \forall i\in\{0,1\}, \param\in\stateSpaceParam\,.
\end{align*}
The choice of this particular uncertainty model can be motivated by the following example.
Consider the linear model
\begin{align*}
    \mathbf  X = \Theta + \mathbf V\,,
\end{align*}
where $\Theta$ is some random quantity of interest and $\mathbf V$ is random noise.
Since the noise distribution is often subject to uncertainty, i.e., $P(\mathbf v\given\theta)$ is not known completely, deploying an uncertainty model on $P(\obs\given\param)$ is a natural choice.

In order to keep the notation compact, the following short notation is introduced $p_{i,\param[i]} := p(\obs\given\Hyp,\param)$. Furthermore, constraints of the form $p_{i,\param[i]} \in \mathcal{P}_{i,\param}$ have to hold for all $i\in\{0,1\}$ and all $\param\in\stateSpaceParam$.

\subsection{Fundamentals of Joint Detection and Estimation}
In \gls{jde}, the aim is to infer the data generating hypothesis $\Hyp$, $i\in\{0,1\}$, as well as the underlying parameter $\param$, $i\in\{0,1\}$.
Treating both problems, the one of detection and the one of estimation separately, does not necessarily result in an overall optimal performance \cite{moustakides2012joint}.
Therefore, both problems have to be solved jointly.

A procedure that solves the problem of \gls{jde} is characterized by a policy $\pi$ that comprises of one decision rule $\dec$ and two estimators $\est[0]$ and $\est[1]$.
In this work a \emph{randomized} decision rule $\dec: \stateSpaceObs \rightarrow [0,1]$ that maps the observations to the probability of accepting the alternative is considered.

In addition to the decision rule, the policy is equipped with a set of estimators, one for each hypothesis.
More formally, the estimator for the parameter under hypothesis $\Hyp$ is defined as $\est:\stateSpaceObs\rightarrow\stateSpaceParam$, $i\in\{0,1\}$.
Hence, they map the observations to the respective state space of the parameter.

The set of all feasible decision rules $\Delta_\delta$ is convex and compact.
Additionally, $\Delta_i$, the set of feasible estimators under hypothesis $\Hyp$  is convex and compact, which directly follows from the definition of $\stateSpaceParam[]$.
From that, it follows that the set of feasible policies $\Pi$ is also convex and compact.

Next, performance measures for \gls{jde} are to be discussed.
For the outcome of the decision, the probabilities of falsely rejecting a hypothesis
\begin{align}
    \begin{split}\label{eq:def_det_err}
        \detErr[0] & = \E[\dec\given\Hyp[0]] = \iint\!\dec p(\obs\given\Hyp[0],\param[0]) p(\param[0]\given\Hyp[0]) \dInt\obs\dInt\param[0]\,,\\
        \detErr[1] & = \E[1-\dec\given\Hyp[1]] = \iint\!(1-\dec) p(\obs\given\Hyp[1],\param[1]) p(\param[1]\given\Hyp[1]) \dInt\obs\dInt\param[1]
    \end{split}
\end{align}
are used as performance measures.
To measure the performance of the estimators, the \gls{mse} is used.
In the case a wrong decision is made, we set the estimation error to zero\footnote{Defining the estimation error in case of a wrong decision is often neither meaningful nor practical as, for example, the random parameter under $\Hyp[1]$ might not even exist under the null hypothesis.}.
Mathematically, the estimation error levels under bot hypotheses can be defined as
\begin{align}
    \begin{split}\label{eq:def_est_err}
         \estErr[0] & = \E\Bigl[(1-\delta)(\est[0]-\param[0])^2\given\Hyp[0]\Bigr] \\
                    & = \iint (1-\delta) (\est[0] - \param[0])^2  p(\obs\given\Hyp[0],\param[0]) p(\param[0]\given\Hyp[0]) \dInt\obs\dInt\param[0]\,,\\
    \estErr[1] & = \E\Bigl[\delta(\est[1]-\param[1])^2\given\Hyp[1]\Bigr] \\
               &= \iint \delta (\est[1] - \param[1])^2  p(\obs\given\Hyp[1],\param[1]) p(\param[1]\given\Hyp[1]) \dInt\obs\dInt\param[1]\,.
    \end{split}
\end{align}
Note that the performance measures in \cref{eq:def_det_err} and \cref{eq:def_est_err} respectively depend on the policy $\pi$ as well as on the conditional densities $p(\obs\given\Hyp[0],\param[0])$ and $p(\obs\given\Hyp[1],\param[1])$.
To keep the upcoming notation compact, these dependencies are only mentioned explicitly if they are relevant within the context.

\subsection{Optimal Joint Detection and Estimation}
A variety of problem formulations exist in \gls{jde}.
In this work, we focus on two different problem formulations.
In the first one, referred to as \emph{Bayesian formulation} in this work, one aims to find a policy that minimizes a weighted sum of detection and estimation errors.
Measuring the inference quality by a linear combination of the individual performance measures is well motivated from a decision theoretic point of view and has been used in the literature \cite{li2007,li2007a,yilmaz2014sequential,yilmaz2015sequential,yilmaz2016sequential}.

Mathematically, the search for the optimal policy under this formulation can be written as the following optimization problem
\begin{align} \label{eq:non_rob_prob_Bayes}
    \tag{P1}
    & \min_{\policy}\; \BayesCost(\policy)\,,
\end{align}
with
\begin{align}\label{eq:objBayes}
 \BayesCost(\policy) = \sum_{i=0}^1P(\Hyp)\left(\detCost[i]\detErr[i](\policy) + \estCost[i]\estErr(\policy)\right)\,.
\end{align}
In \cref{eq:objBayes}, $\detCost, \estCost$, $i\in\{0,1\}$, are some fixed, non-negative cost coefficients.
Although the prior probabilities $p(\Hyp[0])$ and $p(\Hyp[1])$ are fixed, and, hence, could be incorporated into the cost coefficients, they are added to end up with a more compact notation later. 

In the second formulation, referred to as the \emph{Neyman-Pearson-like formulation} or \emph{NP-like formulation}, one aims to find a policy that minimizes a cost function reflecting the estimation errors, while constraining the error probabilities to not exceed nominal levels.
Such formulations have been used, for example, in \cite{moustakides2012joint,jajamovich2012,li2015joint,li2016}.
In the literature, there exist different ways to model the estimation inaccuracy, such as the maximum of the \glspl{mse} under $\Hyp[0]$ and $\Hyp[1]$ \cite{jajamovich2012} or as a weighted sum of both \glspl{mse} \cite{li2015joint,li2016}.
In this work, we focus on the latter.

Mathematically, it can be written as the following optimization problem
\begin{align} \label{eq:non_rob_prob_NP}
    \tag{P2}
    \begin{split}
    & \min_{\policy}\; \NPcost(\policy)\,, \\
    \text{s.t.} \quad & \detErr[0](\policy) \leq \detErrMax[0]\,, \quad \detErr[1](\policy) \leq \detErrMax[1]\,,
    \end{split}
\end{align}
with
\begin{align*}
 \NPcost(\policy) = P(\Hyp[0])\estCost[0]\estErr[0](\policy) + P(\Hyp[1])\estCost[1]\estErr[1](\policy)\,.
\end{align*}
Note that for \cref{eq:non_rob_prob_NP} to have a solution, it has to hold that $\detErrMax[1] > \detErrMax[1]_\text{NP}$, where $\detErrMax[1]_\text{NP}$ is the Type-II error probability of the \gls{NP}-test with nominal Type-I error level $\detErrMax[0]$ \cite{moustakides2012joint}.

\subsection{Minimax Optimal Joint Detection and Estimation}
After the problem formulation under full model knowledge has been provided, the counterparts of \cref{eq:non_rob_prob_Bayes} and \cref{eq:non_rob_prob_NP} under model uncertainty can be introduced.

The minimax version of \cref{eq:non_rob_prob_Bayes} can be described as follows.
We wish to find a policy that minimizes the maximum weighted sum of detection and estimation error levels.
Mathematically, this can be written as the following optimization problem
\begin{align} \label{eq:rob_prob_Bayes}
    \tag{P$1^\ast$}
    & \min_{\policy}\, \max_{P_{i,\param[i]}\in\mathcal{P}_{i,\param}}\; \BayesCost(\policy, p_{0,\param[0]}, p_{1,\param[1]})\,,
\end{align}
where the constraints for the maximum operator have to hold for all $i\in\{0,1\}$ and all $\param\in\stateSpaceParam$.
As a maximization over the set of feasible densities has to be performed, the dependency of the objective on the densities is stated explicitly.
The second and the third argument of $\BayesCost(\pi,\cdot,\cdot)$ have to be read as the collection of $P(\obs\given\Hyp[0],\param[0])$ for all $\param[0]\in\stateSpaceParam[0]$ and $P(\obs\given\Hyp[1],\param[1])$ for all $\param[1]\in\stateSpaceParam[1]$, respectively.

For the \emph{NP-like formulation}, see \cref{eq:non_rob_prob_NP}, the minimax formulation reads as follows
\begin{align} \label{eq:rob_prob_NP}
    \tag{P$2^\ast$}
    \begin{split}
    & \min_{\policy}\; \max_{\substack{P_{i,\param[i]}^\text{D}\in\mathcal{P}_{i,\param}\\
    P_{i,\param[i]}^\text{E}\in\mathcal{P}_{i,\param}}}
    \NPcost(\policy, p_{0,\param[0]}^\text{E}, p_{1,\param[1]}^\text{E})\,, \\
    \text{s.t.} \quad & \max_{\substack{P_{i,\param[i]}^\text{D}\in\mathcal{P}_{i,\param}\\
    P_{i,\param[i]}^\text{E}\in\mathcal{P}_{i,\param}}}\detErr[0](\policy, p_{0,\param[0]}^\text{D}) \leq \detErrMax[0]\,, \\
                      & \max_{\substack{P_{i,\param[i]}^\text{D}\in\mathcal{P}_{i,\param}\\
    P_{i,\param[i]}^\text{E}\in\mathcal{P}_{i,\param}}}\detErr[1](\policy, p_{1,\param[1]}^\text{D}) \leq \detErrMax[1]\,,
    \end{split}
\end{align}
where the constraints for the maximum operators have to hold for all $i\in\{0,1\}$ and all $\param\in\stateSpaceParam$.
The arguments for the function above have to be read as a family of $p(\obs\given\Hyp[i],\param[i])$ for all $\param[i]\in\stateSpaceParam[i]$, $i\in\{0,1\}$.
In the \emph{NP-like formulation}, the objective and the constraint functions have to be maximized separately.
Therefore, one has to maximize over four set of densities, two for maximizing the objective and two for maximizing the constraint functions.

     \section{Minimax Solution under the Bayesian Formulation}\label{sec:Bayes}
In this section, the minimax solution under the Bayesian formulation, i.e., the solution of \cref{eq:rob_prob_Bayes}, is investigated.
Before the characterization of the solution of \cref{eq:rob_prob_Bayes} can be presented, the applicability of Sion's minimax theorem has to be verified.
The objective in \cref{eq:rob_prob_Bayes} is a linear combination of the four different performance measures.
The error probabilities are linear in the decision rule as well as in the densities $p(\obs\given\Hyp,\param)$, $i\in\{0,1\}$.
The \glspl{mse} on the other hand are linear in the decision rule, quadratic in the estimators and linear in the densities $p(\obs\given\Hyp,\param)$, $i\in\{0,1\}$.
This implies that the objective in \cref{eq:rob_prob_Bayes} is continuous in all arguments, it is convex in the policy $\pi$ and concave in the densities  $p(\obs\given\Hyp,\param)$, $i\in\{0,1\}$.
Since the set of feasible policies is compact, Sion's minimax theorem as stated in \cref{theo:minimax} is applicable here.
This implies that finding a solution of \cref{eq:rob_prob_Bayes} is equivalent to first deriving an optimal policy for a given, but arbitrary set of distributions and subsequently finding the \glspl{lfd}.

First, the optimal solution for a given set of distributions is reviewed.
Next, some important properties of the cost functions are presented.
Finally, a characterization of the \glspl{lfd} and the minimax optimal solution is given.

\subsection{Optimal Policy}
In what follows, the optimal policy is presented and briefly discussed.
The derivation is omitted due to space constraints, but it is analog to, e.g., \cite{reinhard2018}.
For easier reference, the optimal policy is fixed in the following theorem.
\begin{theorem}\label{theo:optPolicyBayes}
 Let $\detCost, \estCost\geq0$ be fixed cost coefficients and let the densities $p_{i,\param[i]}$ for $i\in\{0,1\}$ and $\param\in\stateSpaceParam$ be given.
 Then, the optimal policy that solves \cref{eq:non_rob_prob_Bayes} is given by
    \begin{align*}
            \decOpt(\obs) & = \begin{cases}
                                    0, & D_0^\text{B}(\obs) < D^\text{B}_1(\obs)\,, \\
                                    \in[0,1] & D^\text{B}_0(\obs) = D^\text{B}_1(\obs)\,, \\
                                    1, & D^\text{B}_0(\obs) > D^\text{B}_1(\obs)\,,
                                \end{cases}\\
            \estOpt(\obs) & = \E_{p_{i,\param[i]}}[\paramRV\given\Hyp,\obs]\,,
    \end{align*}
    with
    \begin{align*}
        D^\text{B}_i(\obs) = \detCost[1-i] p(\Hyp[1-i] \given \obs) + \estCost p(\Hyp \given \obs) \Var_{p_{i,\param[i]}}[\paramRV\given\Hyp,\obs]\,.
    \end{align*}
    It further holds that
    \begin{align*}
        \BayesCost(\policyOpt, p_{0,\param[0]},p_{1,\param[1]}) = \int \min\Bigl\{D_0^\text{B}(\obs), D_1^\text{B}(\obs)\Bigr\}p(\obs)\dInt\obs\,.
    \end{align*}
\end{theorem}
The optimal estimator under $\Hyp$ as stated in \cref{theo:optPolicyBayes} is the minimum \gls{mse} estimator.
The decision rule accepts $\Hyp[0]$ if the cost for accepting $\Hyp[0]$, i.e., $D^\text{B}_{0}(\obs)$ is smaller than the cost for accepting $\Hyp[1]$, i.e., $D^\text{B}_{1}(\obs)$.
The cost of accepting a particular hypothesis comprises of one term reflecting the certainty about the true hypothesis, whereas a second term reflects the uncertainty about the underlying parameter when this hypothesis is accepted.
Here, the coupling of detection and estimation can be seen as the decision takes detection and estimation accuracy into account.

\subsection{Properties of the Cost Function}\label{sec:prop_cost_Bayes}
At first, the optimal objective from \cref{theo:optPolicyBayes} is rewritten as a function of scalar arguments.
This is essential in order to establish a connection between the current problem and $f$-similarities at a later stage.
Let $\bar{s}_{0,\param[0]} = (s_{\param[0]}^0)_{\param[0]\in\stateSpaceParam[0]}$ and $\bar{s}_{1,\param[1]}=(s_{\param[1]}^1)_{\param[1]\in\stateSpaceParam[1]}$ denote two collections of non-negative scalars that depend on the parameters $\param[0]$ and $\param[1]$, respectively.
First, we define the cost for accepting $\Hyp[i]$ in terms of the collections $\bar{s}_{0,\param[0]}$ and $\bar{s}_{1,\param[0]}$
\begin{align}
\breve D^\text{B}_i(\bar{s}_{0,\param[0]}, \bar{s}_{1,\param[1]}) \notag\\
& = \detCost[1-i] P(\Hyp[1-i])\int s_{\param[1-i]}^{1-i} p(\param[1-i]\given\Hyp[1-i])\dInt\param[1-i] \notag\\
 & \quad + \estCost P(\Hyp) \int \param^2 s_{\param[i]}^{i} p(\param\given\Hyp) \dInt \param \label{eq:BayesCostHypScalarArg}\\
& \quad- \estCost P(\Hyp)\frac{\left( \int \param^2 s_{\param[i]}^{i}p(\param\given\Hyp) \dInt \param \right)^2}{\int s_{\param[i]}^{i}p(\param\given\Hyp)\dInt\param}\,\notag
\end{align}
and a function
\begin{align}\label{eq:BayesCostScalarArg}
    \breve \rho^\text{B}(\bar{s}_{0,\param[0]},& \bar{s}_{1,\param[1]}) = \min\biggl\{
            \breve D^\text{B}_i(\bar{s}_{0,\param[0]}, \bar{s}_{1,\param[1]}) \,,\, 
            \breve D^\text{B}_i(\bar{s}_{0,\param[0]}, \bar{s}_{1,\param[1]})
        \biggr\}\,.
\end{align}
Contrary to the cost functions provided in \cref{theo:optPolicyBayes}, neither the one introduced in \cref{eq:BayesCostHypScalarArg}, nor the one introduced in \cref{eq:BayesCostScalarArg} depend on the observations directly.
Let $\bar{p}_{0,\param[0]} = (p_{0,\param[0]})_{\param[0]\in\stateSpaceParam[0]}$ and $\bar{p}_{1,\param[1]}=(p_{1,\param[1]})_{\param[1]\in\stateSpaceParam[1]}$ denote two collections of \glspl{pdf} that depend on the random parameters $\param[0]$ and $\param[1]$, respectively.
Then, it can be seen that
\begin{align*}
    \BayesCost(\policyOpt, p_{0,\param[0]},p_{1,\param[1]}) = \int \breve \rho^\text{B}(\bar{p}_{0,\param[0]},& \bar{p}_{1,\param[1]}) \dInt\obs\,.
\end{align*}
Though the cost functions in \cref{eq:BayesCostHypScalarArg} and \cref{eq:BayesCostScalarArg} do not depend on the observations directly, the dependence occurs indirectly through the replacement of the collections of scalar by the collection of of conditional \glspl{pdf}.

\begin{theorem}\label{cor:BayesCostScalarArgProp}
    The function in \cref{eq:BayesCostScalarArg} is continuous, concave and homogenous in the arguments $\bar{s}_{0,\param[0]}$ and $\bar{s}_{1,\param[1]}$.
\end{theorem}
A proof is given in \cref{proof:propObjBayes}.
\cref{cor:BayesCostScalarArgProp} implies that \cref{eq:rob_prob_Bayes} admits a solution.
Moreover, according to \cref{cor:BayesCostScalarArgProp}, the function in \cref{eq:BayesCostScalarArg} satisfies all conditions for $f$-similarities, except that its arguments is a collection of scalar variables with possibly an infinite number of elements.
Hence, it could be interpreted as an infinite-dimensional $f$-similarity.
For discrete parameter spaces $\stateSpaceParam[0]$ and $\stateSpaceParam[1]$, it reduces to the finite-dimensional $f$-similarity known from the literature.

In practice, however, the continuous parameter spaces have to be discretized such that the problem reduces to the search of a finite number of densities.
Let $\discrParam[0]{n}$, $n\in\{1,\ldots,N_0\}$, and $\discrParam[1]{n}$, $n\in\{1,\ldots,N_1\}$, denote the grid points of $\stateSpaceParam[0]$ and $\stateSpaceParam[1]$, respectively.
Furthermore, let $\boldsymbol{s}_{i,\param}=\left[s_{i,\param^{[1]}},\ldots,s_{i, \param^{[N_i]}}\right]$ for $i\in\{0,1\}$.
That is, instead of considering collections $\bar{s}_{0,\param[0]}$ and $\bar{s}_{1,\param[1]}$ that may have an infinite number of elements, we consider finite dimensional vectors, whose elements depend on the discretized parameters $\discrParam[i]{n}$, $n\in\{1,\ldots,N_i\}$, $i\in\{0,1\}$.
Now, with the definitions in \cref{eq:discrDefBayes}, the function stated in \cref{eq:BayesCostHypScalarArg} can be further reduced to
\begin{align}\tilde D_i^\text{B}(\boldsymbol s_{0, \param[0]}, \boldsymbol s_{1,\param[1]}) & = \estCost P(\Hyp) \boldsymbol a_i^\top \boldsymbol s_{i,\param} - \estCost P(\Hyp)\dfrac{\Bigl( \boldsymbol b_i^\top \boldsymbol s_{i,\param} \Bigr)^2}{\boldsymbol c_i^\top \boldsymbol s_{i,\param}} \notag\\
 & \quad + \detCost[1-i] P(\Hyp[1-i]) \boldsymbol c_{1-i}^\top \boldsymbol s_{1-i, \param[1-i]}\,,\;\, i\in\{0,1\}\,,    \label{eq:discrDBayes}
\end{align}
with the counterpart of \cref{eq:BayesCostScalarArg} given by
\begin{align}\label{eq:objFctBayesDiscr}
    \begin{split}
    \rho^{B}(&\boldsymbol s_{0, \param[0]}, \boldsymbol s_{1,\param[1]}) \\
    & =  \min\left\{\tilde D_0^\text{B}(\boldsymbol s_{0, \param[0]}, \boldsymbol s_{1,\param[1]})\,,\,\tilde D_1^\text{B}(\boldsymbol s_{0, \param[0]}, \boldsymbol s_{1,\param[1]})\right\}\,.
    \end{split}
\end{align}
\begin{figure*}[t!]
\begin{align}\label{eq:discrDefBayes}
\bigl[ \boldsymbol{a}_i \bigr]_n = \Bigl(\discrParam{n}\Bigr)^2p\Bigl(\discrParam{n}\Given\Hyp\Bigr)\Delta\param\,,\quad
\bigl[ \boldsymbol{b}_i \bigr]_n = \discrParam{n}p\Bigl(\discrParam{n}\Given\Hyp\Bigr)\Delta\param \,,\quad
\bigl[ \boldsymbol{c}_i \bigr]_n = p\Bigl(\discrParam{n}\Given\Hyp\Bigr)\Delta\param
\end{align}
    \hrulefill
    \vspace*{4pt}  
\end{figure*}

Finally, we are able to show that the \glspl{lfd} are given by the maximizer of an $f$-similarity.
This is stated in the next corollary.
Let $\boldsymbol p_{i,\param} = \left[p_{i,\param[i]^{[1]}}, \ldots, p_{i,\param[i]^{[N_i]}}\right]$ be vectors of conditional \glspl{pdf}, $\boldsymbol P_{i,\param}$ the corresponding probability distribution functions and let $\mathbb{P}_i = \bigtimes_{n=1}^{N_i} \mathcal{P}_{i,\param[i]^{[n]}}$ the corresponding uncertainty sets for $i\in\{0,1\}$.
\begin{corollary}\label{cor:lfd_Bayes}
    The \glspl{lfd} that solve \cref{eq:rob_prob_Bayes} after discretizing the parameter spaces are given by
    \begin{align*}
\boldsymbol Q_{0,\param[0]}, \boldsymbol Q_{1, \param[1]} = \argmax_{{\boldsymbol P_0 \in \mathbb{P}_0\,,\,\boldsymbol P_1 \in \mathbb{P}_1}}\;
D_{\rho^\text{B}}(\boldsymbol P_{0,\param[0]}, \boldsymbol P_{1,\param[1]})
    \,,
    \end{align*}
    where $D_{\rho^\text{B}}(\cdot)$ is the $f$-similarity induced by $\rho^\text{B}(\cdot)$ as defined in \cref{eq:objFctBayesDiscr}.
\end{corollary}
    The fact that $\rho^{\text{B}}(\cdot)$ induces a valid $f$-similarity follows directly from \cref{cor:BayesCostScalarArgProp} and the finite dimensionality of its arguments.

For given $\detCost, \estCost\geq0$, $i\in\{0,1\}$, the minimax optimal solution is given by a policy as stated in \cref{theo:optPolicyBayes}, where the nominal distributions are replaced by the distributions found as in \cref{cor:lfd_Bayes}.

     \section{Minimax Solution under the NP-like Formulation}\label{sec:NP}
In this section, the minimax solution under an \gls{NP}-like formulation, i.e., the solution of \cref{eq:rob_prob_NP}, is investigated.
As a first step, the constrained problem in \cref{eq:rob_prob_NP} has to be converted to an unconstrained one.
The weighted cost function is given by
\begin{align}\NPcostUnconstr(\policy,p^\text{D}_{0,\param[0]}, p^\text{D}_{1,\param[1]}, p^\text{E}_{0,\param[0]}, p^\text{E}_{1,\param[1]})
&= \sum_{i=0}^1 P(\Hyp)\biggl(\detCost[i]\detErr[i](\policy,p^\text{D}_{i,\param[i]}) \notag\\
&\quad + \estCost[i]\estErr[i](\policy,p^\text{E}_{i,\param[i]})\biggr)\,. \label{eq:NP_obj_unconstr}
\end{align}
where $\detCost\geq0$ are assumed to be fixed for now.
The choice of $\detCost$ is addressed later in this section.
With the same line of arguments as in \cref{sec:Bayes}, we can state that Sion's minimax theorem is applicable to the objective in \cref{eq:NP_obj_unconstr}.
Hence, we can proceed as in \cref{sec:Bayes} and first provide the optimal policy for given coefficients and densities, and subsequently characterize the \glspl{lfd}.

\subsection{Optimal Policy}\label{sec:opt_pol_NP}
The optimal policy that minimizes the cost function stated in \cref{eq:NP_obj_unconstr} can be obtained similarly as for the \emph{Bayesian formulation}.
For the sake of clarity and for easier referencing, it is summarized in the following theorem.
\begin{theorem}\label{theo:optPolicyNP}
 For given non-negative coefficients $\detCost, \estCost$, $i\in\{0,1\}$, and fixed densities $p^\text{D}_{i,\param[i]}, p^\text{E}_{i,\param[i]}$, $i\in\{0,1\}, \param\in\stateSpaceParam$, the policy $\policyOpt = (\decOpt, \estOpt[0], \estOpt[1])$ that minimizes \cref{eq:NP_obj_unconstr} is given by
 \begin{align*}
        \decOpt(\obs) & = \begin{cases}
        0 & D_0^\text{NP}(\obs) < D_1^\text{NP}(\obs)\,, \\
        \kappa\in[0,1] & D_0^\text{NP}(\obs) = D_1^\text{NP}(\obs)\,, \\
        1 & D_0^\text{NP}(\obs) > D_1^\text{NP}(\obs)\,,
        \end{cases}\\
        \estOpt(\obs) & = \E_{p^\text{E}_{i,\param[i]}}[\paramRV\given\Hyp,\obs]\,,
 \end{align*}
with
\begin{align*}
D^\text{NP}_i(\obs) & = P(\Hyp[1-i]) \detCost[1-i]  \detDens[1-i](\obs\given\Hyp[1-i]) \\
& \quad + P(\Hyp[i])\estCost[i]  \Var_{p^\text{E}_{i,\param[i]}}[\paramRV\given\Hyp[i],\obs] \estDens[i](\obs\given\Hyp[i])\,.
\end{align*}
It further holds that
\begin{align*}
\NPcostUnconstr(&\policyOpt, p^\text{D}_{0,\param[0]}, p^\text{D}_{1,\param[1]}, p^\text{E}_{0,\param[0]}, p^\text{E}_{1,\param[1]}) \\
&= \int \min\Bigl\{D_0^\text{NP}(\obs), D_1^\text{NP}(\obs)\Bigr\}\dInt\obs\,.
\end{align*}
\end{theorem}
The cost minimizing policy stated above looks similar to the one stated in \cref{sec:Bayes}.
However, in the Bayesian case the optimal policy only depends on two sets of distributions, whereas the policy given in the theorem above depends on four sets of distributions.
This is caused by the fact that the detection and estimation error measures have to be maximized separately in the sequel.
This fact also affects the structure of te cost for accepting hypothesis $\Hyp$, i.e., $D_i^\text{NP}(\obs)$.

\subsection{Properties of the Cost Function}
Here, we proceed similarly to \cref{sec:prop_cost_Bayes} and first introduce a function that depends on four collections of scalars, where each collection may have an infinite number of elements.
Let $\bar{s}^\text{D}_{0,\param[0]} = (s_{\param[0]}^{\text{D},0})_{\param[0]\in\stateSpaceParam[0]}$,
$\bar{s}^\text{D}_{1,\param[1]} = (s_{\param[1]}^{\text{D},1})_{\param[1]\in\stateSpaceParam[1]}$, $\bar{s}^\text{E}_{0,\param[0]} = (s_{\param[0]}^{\text{E},0})_{\param[0]\in\stateSpaceParam[0]}$ and $\bar{s}^\text{E}_{1,\param[1]} = (s_{\param[1]}^{\text{E},1})_{\param[1]\in\stateSpaceParam[1]}$ denote four collections of non-negative scalars whose elements depend on the parameters $\param[0]$ and $\param[1]$, respectively.
Then, we can rewrite the cost for accepting $\Hyp$ as follows
\begin{align}
    \begin{split} \label{eq:NPCostHypScalarArg}
 \breve D^\text{NP}_i(&\bar{s}^\text{D}_{0,\param[0]}, \bar{s}^\text{D}_{1,\param[1]}, \bar{s}^\text{E}_{0,\param[0]}, \bar{s}^\text{E}_{1,\param[1]})  \\
& = \detCost[1-i] P(\Hyp[1-i])\int s_{\param[1-i]}^{\text{D},1-i} p(\param[1-i]\given\Hyp[1-i])\dInt\param[1-i]
\\
& \quad + \estCost P(\Hyp)
            \int \param^2 s_{\param[i]}^{\text{E},i} p(\param\given\Hyp) \dInt \param
\\
& \quad- \estCost P(\Hyp)\frac{\left( \int \param^2 s_{\param[i]}^{\text{E},i}p(\param\given\Hyp) \dInt \param \right)^2}{\int s_{\param[i]}^{\text{E},i}p(\param\given\Hyp)\dInt\param}\,,
    \end{split}
\end{align}
and a function
\begin{align}\label{eq:NPCostScalarArg}
        \begin{split}
        \breve\rho^\text{NP}(&\bar{s}^\text{D}_{0,\param[0]}, \bar{s}^\text{D}_{1,\param[1]}, \bar{s}^\text{E}_{0,\param[0]}, \bar{s}^\text{E}_{1,\param[1]}) \\
        &= \min\Bigl\{\breve D_0^\text{NP}(\bar{s}^\text{D}_{0,\param[0]}, \bar{s}^\text{D}_{1,\param[1]}, \bar{s}^\text{E}_{0,\param[0]}, \bar{s}^\text{E}_{1,\param[1]}) , \\
        & \quad \breve D_1^\text{NP}(\bar{s}^\text{D}_{0,\param[0]}, \bar{s}^\text{D}_{1,\param[1]}, \bar{s}^\text{E}_{0,\param[0]}, \bar{s}^\text{E}_{1,\param[1]}) \Bigr\}\,.
        \end{split}
\end{align}
which do, similar to their counterpart in \cref{sec:prop_cost_Bayes}, not directly depend on the observations.
\begin{theorem}\label{cor:NPCostScalarArgProp}
    The function in \cref{eq:NPCostScalarArg} is continuous, concave and homogenous in the arguments $\bar{s}^\text{D}_{0,\param[0]}, \bar{s}^\text{D}_{1,\param[1]}, \bar{s}^\text{E}_{0,\param[0]}$ and $\bar{s}^\text{E}_{1,\param[1]}$.
\end{theorem}
The proof can be carried out using the same line of arguments as that in \cref{proof:propObjBayes}.

Next, by discretizing the continuous parameter spaces $\stateSpaceParam[0]$ and $\stateSpaceParam[1]$ by $\discrParam[0]{n}$, $n\in\{1,\ldots,N_0\}$, and $\discrParam[1]{n}$, $n\in\{1,\ldots,N_1\}$, we can define $\boldsymbol{s}_{i}^\text{D}=[s^\text{D}_{i,1},\ldots,s^\text{D}_{i, N_i}]$ and $\boldsymbol{s}^\text{E}_{i}=[s^\text{E}_{i,1},\ldots,s^\text{E}_{i, N_i}]$ for $i\in\{0,1\}$.

After that, the cost for accepting hypothesis $\Hyp$ becomes
\begin{align}\label{eq:discrDNP}
    \begin{split}
 \tilde D_i^\text{NP}(&\boldsymbol s_{0,\param[0]}^\text{D}, \boldsymbol s_{1,\param[1]}^\text{D}, \boldsymbol s_{0,\param[0]}^\text{E}, \boldsymbol s_{1,\param[1]}^\text{E}) \\
 & = \estCost P(\Hyp) \boldsymbol a_i^\top \boldsymbol s^\text{E}_{i,\param[i]} 
   - \estCost P(\Hyp)\frac{\left( \boldsymbol b_i^\top \boldsymbol s^\text{E}_{i,\param[i]} \right)^2}{\boldsymbol c_i^\top \boldsymbol s^\text{E}_{i,\param[i]}}\\
   & \quad + \detCost[1-i] P(\Hyp[1-i]) \boldsymbol c_{1-i}^\top \boldsymbol s_{1-i,\param[1-i]}^\text{D}\,.
    \end{split}
\end{align}
with the counterpart of \cref{eq:NPCostScalarArg} given by
\begin{align}\label{eq:objFctNPDiscr}
    \begin{split}
    \rho^\text{NP}(&\boldsymbol s_{0,\param[0]}^\text{D}, \boldsymbol s_{1,\param[1]}^\text{D}, \boldsymbol s_{0,\param[0]}^\text{E}, \boldsymbol s_{1,\param[1]}^\text{E}) \\
    & =\min\Bigl\{\tilde D_0^\text{NP}(\boldsymbol s_{0,\param[0]}^\text{D}, \boldsymbol s_{1,\param[1]}^\text{D}, \boldsymbol s_{0,\param[0]}^\text{E}, \boldsymbol s_{1,\param[1]}^\text{E})\,,\,\\
    & \quad \tilde D_1^\text{NP}(\boldsymbol s_{0,\param[0]}^\text{D},\boldsymbol s_{1,\param[1]}^\text{D}, \boldsymbol s_{0,\param[0]}^\text{E}, \boldsymbol s_{1,\param[1]}^\text{E})\Bigr\}\,.
    \end{split}
\end{align}
The remaining symbols are defined in \cref{eq:discrDefBayes}.
Finally, we are able to show that the densities that maximize the weighted cost function are given by the maximizer of an $f$-similarity.
This is stated in the next corollary.
Let $\boldsymbol p_i^\text{D} = \left[p^\text{D}_{i,\param[i]^{[1]}}, \ldots, p^\text{D}_{i,\param[i]^{[N_i]}}\right]$ and $\boldsymbol p_i^\text{E} = \left[p^\text{E}_{i,\param[i]^{[1]}}, \ldots, p^\text{E}_{i,\param[i]^{[N_i]}}\right]$ be vectors of conditional \glspl{pdf}, $\boldsymbol P_i^\text{D}$, $\boldsymbol P_i^\text{E}$ the corresponding probability distribution functions,  and let the corresponding uncertainty sets be $\mathbb{P}_i = \bigtimes_{n=1}^{N_i} \mathcal{P}_{i,\param[i]^{[n]}}$ for $i\in\{0,1\}$.
To shorten the upcoming notation, we will denote the collection of all \glspl{lfd} as 
$\boldsymbol Q = (\boldsymbol Q_{0,\param[0]}^\text{D}, \boldsymbol Q_{1,\param[1]}^\text{D}, \boldsymbol Q_{0,\param[0]}^\text{E}, \boldsymbol Q_{1,\param[1]}^\text{E})$.
\begin{corollary}\label{cor:f-sim_NP}
    For given $\detCost, \estCost\geq0$, $i\in\{0,1\}$, the least favorable distributions that maximize \cref{eq:NP_obj_unconstr} after discretizing the parameter spaces are given by
    \begin{align*}
        \boldsymbol Q =
        \argmax_{\substack{\boldsymbol P_0^\text{D} \in \mathbb{P}_0\,,\;\boldsymbol P_0^\text{E} \in \mathbb{P}_0 \\ \boldsymbol P_1^\text{D} \in \mathbb{P}_1\,,\;\boldsymbol P_1^\text{E} \in \mathbb{P}_1}}
        \; D_{\rho^\text{NP}}(\boldsymbol P_0^\text{D}, \boldsymbol P_1^\text{D}, \boldsymbol P_0^\text{E}, \boldsymbol P_1^\text{E})\,,
    \end{align*}
    where $D_{\rho^\text{NP}}(\cdot)$ is the $f$-similarity induced by $\rho^\text{NP}(\cdot)$ as defined in \cref{eq:objFctNPDiscr}.
\end{corollary}
Note that the distributions in \cref{cor:f-sim_NP} are least favorable for \cref{eq:NP_obj_unconstr}, but not necessarily for \cref{eq:rob_prob_NP}.
\subsection{Design of the Minimax Optimal Procedure}
After the optimal policy for some given set of distributions and the characterization of the distributions maximizing the weighted sum of error measures have been presented, the design of minimax optimal procedures is addressed.

The solution of the minimax problem formulation in \cref{eq:rob_prob_NP} is summarized in the following theorem.
\begin{theorem}\label{theo:minimax_NP}
    Given two non-negative constants $\estCost[0]$ and $\estCost[1]$, two disjoint uncertainty sets $\mathbb{P}_0$ and $\mathbb{P}_1$ corresponding to the discretized parameter spaces and nominal detection error levels $\detErrMax[0]\in(0,0.5)$ and $1>\detErrMax[1]>\detErrMax[1]_\text{NP}>0$, where $\detErrMax[1]_\text{NP}$ is the Type-II error probability of the \gls{NP} test (using $\boldsymbol q^\text{D}_{0,\param[0]}$ and  $\boldsymbol q^\text{D}_{1,\param[1]}$) with Type-I error probability $\detErrMax[0]$.
    Then, a policy as given in \cref{theo:optPolicyNP}
    and the distributions according to \cref{cor:f-sim_NP}
    also solve \cref{eq:rob_prob_NP} for discretized parameter spaces, if it holds that
    \begin{align} \label{eq:detErr_NP}
        \detErr[0](\policyOpt,  \boldsymbol q^\text{D}_{0,\param[0]}) = \detErrMax[0]\,,
         \qquad
         \detErr[1](\policyOpt,  \boldsymbol q^\text{D}_{1,\param[1]}) = \detErrMax[1].
    \end{align}
    The latter is fulfilled if one chooses $\detCost[0]$ and $\detCost[1]$ such that
    \begin{align}\label{eq:optCostCoeff}
        \detCost[0]^\star, \detCost[1]^\star = \argmax_{\detCost[0], \detCost[1]} \NPcostUnconstrDiscr(\allDensLFD; \policyOpt) - \sum_{i=0}^1\detCost p(\Hyp)\detErrMax\,.
    \end{align}
\end{theorem}
A proof is outlined in \cref{proof:minimax_NP}.
\cref{theo:minimax_NP} states that the solution of \cref{eq:rob_prob_NP} is an \emph{optimal} procedure for a set of \emph{least favorable distributions}.
However, the \glspl{lfd} depend on the optimal policy and vice versa.

To design a minimax procedure in the sense of \cref{theo:minimax_NP}, we proceed as follows.
As an initial step, the optimal procedure for \emph{some} densities in $\mathbb{P}_0$ and $\mathbb{P}_1$ is designed.
Subsequently, one iteratively calculates the cost maximizing distributions according to \cref{cor:f-sim_NP}, followed by a search for the optimal cost coefficients.
This process is repeated until convergence.

Note that this iterative design process cannot be interpreted as iteratively designing an optimal scheme and finding a set of \glspl{lfd}.
Once an optimal scheme is designed, one can find a set of most similar, or least favorable, distributions.
However, once the coefficients $\detCost[0]$ and $\detCost[1]$ are updated, i.e., an optimal scheme for this new set of densities is designed, the densities are no longer most similar in the sense of \cref{cor:f-sim_NP}.
Once the iterative design process converges, the distributions are least favorable in the sense of \cref{eq:rob_prob_NP} and the resulting procedure is the minimax optimal solution.
     \section{Discussion}\label{sec:discussion}
In \cref{sec:Bayes,sec:NP}, we have shown that the least favorable, or most similar, distributions that lead to a minimax scheme are defined as the maximizer of some $f$-similarity.
This is in line with classic as well as more recent results in the context of minimax robust hypothesis testing.
As one of the earliest important results, the characterization of the \glspl{lfd} of a binary fixed-sample size test by Huber and Strassen \cite{huber1973minimax} has to be mentioned.
This says that the pair of distributions that simultaneously minimize \emph{all} $f$-divergences are the \glspl{lfd} for a binary hypothesis test for all sample sizes and all thresholds.
However, besides the case of binary hypothesis testing, no such strong conditions exist.
Though the results by Huber and Strassen could be extended to $M$-ary hypothesis testing by replacing $f$-divergences by $f$-dissimilarities, there is no set of distributions that jointly minimizes all $f$-dissimilarities.
Intuitively, this can be explained as in the $M$-ary case where there are multiple sources of error, i.e., more than confusing $\Hyp[0]$ and $\Hyp[1]$.
Hence, one could only find a pair of \glspl{lfd} that maximize a particular sum of weighted error probabilities.
See \cite[Sec. VI]{fauss2021} for a discussion.
Additionally, it has been shown in the context of sequential hypothesis testing that the \glspl{lfd} are defined as the maximizer of some \emph{particular} $f$-similarity \cite{fauss2016,fauss2020minimax}.
However, similar to \cref{sec:NP}, one has to \emph{jointly} find a policy and the \glspl{lfd}.
Hence, one can see that the results presented in this work are in line with existing results in minimax robust hypothesis testing.
Although this paper only addresses \gls{jde} for a binary hypothesis testing problem, there are also multiple sources of errors.
Besides confusing $\Hyp[0]$ and $\Hyp[1]$, also errors during the estimation could occur.
The problem of \gls{jde} can also be seen as an $M$-ary hypothesis test with an infinite number of hypotheses.
Following this interpretation underpins the fact that the minimax optimal solution only minimize a particular $f$-dissimilarity. 
Therefore, it is only natural that a condition as strong as the one derived by Huber and Strassen could not exist. Instead, the set of \glspl{lfd} maximizes a particular $f$-similarity.     \section{Calculation of the Most Similar Distributions}\label{sec:calc_distr}
As shown in \cref{sec:Bayes,sec:NP}, the most similar or least favorable distributions are defined as the maximizer of some $f$-similarity, i.e., of some concave function.
As the uncertainty sets are assumed to be convex, the \glspl{lfd} can, in principle, be found by solving a convex optimization problem.
However, the optimization has to be performed over $N$ \glspl{pdf}, where $N=N_0 + N_1$ in the Bayesian or $N=2\cdot N_0+2\cdot N_1$ in the \gls{NP}-like formulation.
In practice, $N$ could be in the range of hundreds or thousands, and, hence, solving this problem is highly challenging.
As mentioned earlier, we use the band model for implementing the robust procedure.
Therefore, we resort to an adaptation of \cite{fauss2018_tsp}, which is an algorithm tailored to solving problems of the form
\begin{align}\label{eq:impl_prob}
    \min_{\bm P\in\densSet} \; \int f(\allDens) \dInt\obs\,,
\end{align}
where $f(\allDens)$ is a convex function of a vector of $N$ \glspl{pdf}, $\bm P$ and $\densSet$ are the corresponding distributions and band-type uncertainty sets.
In the context of this work, $f(\cdot)$ is replaced by the functions $-\rho^\text{B}(\cdot)$ and $-\rho^\text{NP}(\cdot)$, respectively, so that solving \cref{eq:impl_prob} is equivalent to maximizing the $f$-similarities introduced in \cref{sec:Bayes,sec:NP}, respectively.
The optimal densities are then approximated iteratively, where in each iteration the optimization problem reduces to a one-dimensional line search for a normalization constant.
Using this normalization constant, the most similar distributions are found by projecting the inverse of the objective's subdifferential onto the set of feasible densities.
To increase numerical stability, the algorithm presented in \cite{fauss2018_tsp} comes with a proximal version.

We propose two modifications for the algorithm as described in \cite{fauss2018_tsp}.
First, we replace the Euclidean proximal term by the more general Bregman distance.
Second, we deploy an adaptive weighting scheme to the Bregman distance.

The proximal formulation in \cite{fauss2018_tsp} is generalized by replacing the Euclidean term by a Bregman map $h(\cdot,\cdot)$ \cite{bregman1967relaxation}. 
Now, the new objective of the proximal algorithm can be written as
\begin{align*}\allDensLFD^{(k)} = \min_{\bm P\in\densSet} \; \int f(\allDens) + \eta_k h\Bigl(\allDens, \allDensLFD^{(k-1)}\Bigr)\dInt\obs := \min_{\bm P\in\densSet} \tilde f(\allDens)\,,
\end{align*}
where $\allDensLFD^{(k)}$ denotes the solution at the $k$-th iteration and $\eta_k$ the weight at iteration $k$.

The class of Bregman maps contains various widely used function such as the Euclidean distance or the generalized \gls{KL}-divergence.
In the case where the free variables are valid densities, the latter reduces to the \gls{KL}-divergence.
Since the \gls{KL}-divergence is a widely used statistical distance measure and has some advantages such as reflecting the differences in the tails in an adequate manner, we resort to the \gls{KL}-divergence in this work.
Then, the Bregman map reads
\begin{align*}
    h\Bigl(\allDens, \allDensLFD^{(k-1)}\Bigr) = \sum_{n=1}^N\int p_n(\obs) \log\left(\frac{p_n(\obs)}{q_n^{(k-1)}(\obs)}\right)\dInt \obs\,.
\end{align*}
As mentioned above, \cite{fauss2018_tsp} relies on calculating the inverse sub-differentials of the objective. For any function $\tilde f: \mathbb{R}^N_{\geq 0} \rightarrow \mathbb{R}$, we denote the partial sub-derivatives by $\tilde f_n$, $n\in\{1,\ldots,N\}$. With the generalized \gls{KL}-divergence, the partial sub-derivative becomes
\vspace*{-0.1cm}
\begin{align*}
\tilde f_n(\allDens) = f_n(\allDens) + \eta_k \int \log p_n(\obs)- \log q_n^{(k-1)}(\obs)\dInt \obs\,.
\end{align*}
The partial sub-derivatives of the objective $f(\allDens)$ depend on whether the Bayesian or the \gls{NP}-like formulation is used and are derived in \cref{sec:gradient}.

Instead of using a constant $\eta_k$, we propose an exponentially decaying sequence $\{\eta_k\}_{k\geq0}$.
This should trade-off convergence speed and numerical stability.
More formally, the sequence is defined as
\begin{align*}
    \eta_k = \min\Bigl\{\eta_0 \rho^{k-k_\text{init}}\,,\,\eta_\text{min}\Bigr\}\,,
\end{align*}
where $\eta_0>0$ is the initial weight, $\rho\in(0,1)$ is a decay factor, $\eta_\text{min}$ is the minimum weight and $k_\text{init}$ is the iteration in which the exponential decay starts.
The latter is used to reset the exponential decay to its initial value when running into numerical instability.
We set $k_\text{init}=0$ so that the sequence $\eta_k$ decays exponentially until the process ist reset the first time.
Once the algorithm runs into instability, e.g., the algorithm in \cite{fauss2018_tsp} is not able to find a valid density, $k_\text{init}$ is set to the current iteration number.
The idea behind this exponential weighting scheme is as follows.
Instead of most proximal optimization methods, we interpret $\nabla_k h(\cdot,\cdot)$ as a regularization term.
We further assume that close to the optimal solution, the optimization problem is almost well conditioned, i.e., it does not run into numerical instability.
In early iterations, i.e., when the current solution is far from the optimal one, it requires a lot of regularization to keep the problem well conditioned.
Once the solution moves towards the optimal solution, less regularization is required and too much regularization will require many more iterations until convergence.
To avoid numerical instability, the weight must not fall below $\eta_\text{min}$ and, in case the algorithm still becomes numerically unstable, the exponential decaying is reset.
     \section{Numerical Results}\label{sec:num_results}
In this section, we consider a slightly modified version of the system model in \cref{eq:model}, i.e.,
\begin{align}\label{eq:model_ex1}
    \begin{split}
        \Hyp[0]: & \quad  \obsRV \iid P(\obs\given\Hyp[0])\,,\\
        \Hyp[1]: & \quad  \obsRV\given\param[1] \iid P(\obs\given\Hyp[1],\param[])\,, \quad \paramRV[] \sim P(\param[]\given\Hyp[1])\,.
    \end{split}
\end{align}
In this model, the null hypothesis models a nominal state, whereas the alternative models an abnormal state.
Under the latter, the distribution depends on a parameter of interest.
Though this model is simpler than \cref{eq:model}, it covers a variety of applications \cite{moustakides2012joint}.
In radar for example, the null hypothesis models the \emph{no target} scenario and the alternative the \emph{target present} scenario, where the parameter might be the target location.

In particular, we assume that under the null hypothesis, the data follows a zero-mean Gaussian distribution and under the alternative a Gaussian distribution with location parameter $\paramRV[]$ that is uniformly distributed on the interval $[1, 6]$.
The variance of $\obsRV\given\Hyp[0]$ and $\obsRV\given\Hyp[1],\param[1]$ is set to $\sigma^2=1$ for all $\param[1]\in\stateSpaceParam[1]$

To model the distributional uncertainties, a band model with the following specifications is used
\begin{align*}
 \mathcal{P}_0 & = \biggl\{P: p(x)\geq \underbar{c}\cdot\varphi(x),\;p(x)\leq \bar{c}\cdot\varphi(x)\biggr\}\,, \\
 \mathcal{P}_1^{\param[]} & = \biggl\{P:  p(x)\geq \underbar{c}\cdot\varphi(x-\param[]),\; p(x)\leq \bar{c}\cdot\varphi(x-\param[]) \biggr\}\,,
\end{align*}
where $\varphi(x)$ denotes the \gls{pdf} of the standard normal distribution.
In what follows, we numerically design minimax optimal procedures under the Bayesian and under the \gls{NP}-like formulation for the setup mentioned above.
All numerical results were produced using $10^6$ Monte Carlo runs.

\begin{figure*}[!t]
  \subfloat[\label{fig:lfd_Bayes}Least-favorable densities including corresponding density band under $\mathrm{H}_0$ (\textcolor{my_blue}{{$\pmb -$}}) and under $\mathrm{H}_1$ for different values of $\theta$: $\theta=1.00$ (\textcolor{my_red}{{$\pmb -$}}),
  $\theta=3.50$ (\textcolor{my_green}{{$\pmb -$}}),
  $\theta=6.00$ (\textcolor{my_violet}{{$\pmb -$}}).]{\includegraphics[]{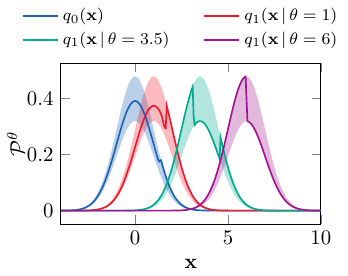}}
  \hfill
 \subfloat[\label{fig:dec_Bayes}Decision rules of the optimal procedure (\textcolor{my_blue}{{$\pmb -$}}) and the minimax optimal procedure (\textcolor{my_red}{{$\pmb -$}}). ]{\includegraphics[]{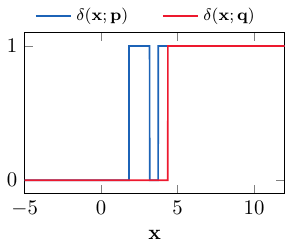}}
 \hfill
 \subfloat[\label{fig:postvar_Bayes}Posterior variance under the nominal (\textcolor{my_blue}{{$\pmb -$}}) and under the least favorable model (\textcolor{my_red}{{$\pmb -$}}).]{\includegraphics{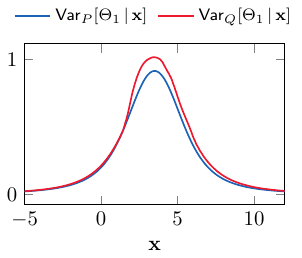}}
 \caption{\label{fig:res_Bayes}Least-favorable densities and resulting decision rule under the Bayesian formulation.}
\end{figure*} 

\subsection{Bayesian Formulation}\label{subsec:num_Bayes}
For the Bayesian setup, we manually set the detection costs to $\detCost[0]=0.75, \detCost[1]=1$ and the estimation cost to $\estCost[1]=1.1$.
Additionally, we chose $\underbar{c}=0.8$ and $\bar{c}=1.2$ for the neighborhood.

In \cref{fig:res_Bayes}, the least-favorable densities, the resulting decision rule and the resulting posterior variance are shown.
Recall, that the \glspl{lfd} maximize a weighted sum of detection and estimation error levels. 
This trade-off between detection and estimation error levels can be seen in \cref{fig:lfd_Bayes}.
From the conditional density for $\param[]=6$, one can see that most of the probability mass is shifted to the left, i.e., towards the density under $\Hyp[0]$.
Besides that, more probability mass is assigned to regions of the state space which lead to a higher posterior variance, see \cref{fig:postvar_Bayes}. 
For $\param[]=3.5$, the conditional distribution leads to a high variance, as most of the probability mass is shifted to the left and to the right.
Additionally, it can be seen that more mass is shifted to the left than to the right, which is due to the fact that it should maximize the variance as well as its similarity to $q_0(\obs)$. 
For $\param[]=1$, the probability mass is shifted to the right, i.e., to a region which has a high posterior variance.
However, this effect is not as pronounced as for $\param[]=6$ since maximizing the similarity between both hypothesis counterweights the effect of variance maximization.
For the density under the null hypothesis, which does not contain a parameter of interest, little mass is shifted to the right, i.e. to the region of $\Hyp[1]$.

For the decision rules depicted in \cref{fig:dec_Bayes}, it can be seen that with increasing value of $\obs$, the optimal decision rule first switches from a decision from $\Hyp[0]$ to $\Hyp[1]$, back to a decision in favor of $\Hyp[0]$ and finally, for $\obs>3.75$ to a decision in favor of $\Hyp[1]$.
This is caused by the effect that there exist a small region of the state space in which the posterior variance overweighs the probability of a wrong decision.
For the decision rule using the \glspl{lfd}, the effect is not visible at all.
However, this does  not mean that there is no coupling between detection and estimation.
Due to the definition of the decision rule, it takes always the posterior variance, as a measure of estimation uncertainty, into account.

For the posterior variance as shown in \cref{fig:postvar_Bayes}, it can be seen that the \glspl{lfd} maximize it almost over the entire state space.

\begin{table}[!t]
    \centering
    \caption{Bayesian formulation: results of Monte Carlo simulations.}
    \begin{tabular}{c|c|c|c|c}
&  & \multicolumn{2}{c|}{error probabilities}\\
        & $\BayesCost$ & Type-I & Type-II & \gls{mse}\\        
        \toprule
        opt. with $\allDens$ & $0.417$& $0.033$& $0.305$ & $0.446$\\
        opt. with $\allDensLFD$ & $0.466$& $0.038$& $0.322$ & $0.533$ \\
        minimax with $\allDens$ & $0.427$& $0.000$ & $0.672$ & $0.171$ \\
        minimax with $\allDensLFD$ &$0.446$ & $0.000$ & $0.688$ & $0.190$
    \end{tabular}
    \label{tbl:Bayes_mc}
\end{table}

The results of the Monte Carlo simulations are summarized in \cref{tbl:Bayes_mc}.
Here, one can directly see from the second column, i.e., the Monte Carlo evaluation of $\BayesCost$, that these results are in line with Sion's minimax theorem.
More precisely, the optimal and the minimax optimal procedure lead to lower cost $\BayesCost$ when evaluated under the nominal densities.
Additionally, a shift in the error probabilities can be observed for the minimax procedure, i.e., the minimax never accepts falsely the alternative. At the same time, it falsely accepts the null hypothesis with a probability of more than $60\,\%$.
This is caused by the following fact.
The decision rule is determined by weighted sums of performance measures.
While the measures themselves are subject to maximization, their weights are not touched at all.
As the measures change in a different numerical range during optimization, the decision rule is highly affected, see \cref{fig:dec_Bayes}.

 \begin{figure}[htp!]
    \centering
  \subfloat[\label{fig:lfd_det_np}Least-favorable densities used for calculating the error probabilities including corresponding density band under $\mathrm{H}_0$ (\textcolor{my_blue}{{$\pmb -$}}) and under $\mathrm{H}_1$ for different values of $\theta$: $\theta=1.00$ (\textcolor{my_red}{{$\pmb -$}}),
  $\theta=3.5$ (\textcolor{my_green}{{$\pmb -$}}),
  $\theta=6.00$ (\textcolor{my_violet}{{$\pmb -$}}).]{\includegraphics[]{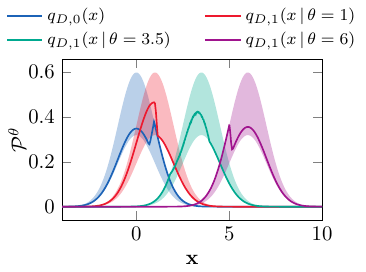}}
  \hfill
  \subfloat[\label{fig:lfd_est_np}Least-favorable densities used for calculating the \gls{mse} including corresponding density band under $\mathrm{H}_1$ for different values of $\theta$: $\theta=1.00$ (\textcolor{my_red}{{$\pmb -$}}),
  $\theta=3.5$ (\textcolor{my_green}{{$\pmb -$}}),
  $\theta=6.00$ (\textcolor{my_violet}{{$\pmb -$}}).]{\includegraphics[]{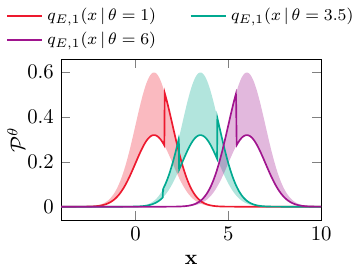}}
  \hfill  
 \subfloat[\label{fig:dec_np}Decision rules of the optimal (\textcolor{my_blue}{{$\pmb -$}}) and the minimax optimal procedure (\textcolor{my_red}{{$\pmb -$}}). ]{\includegraphics[]{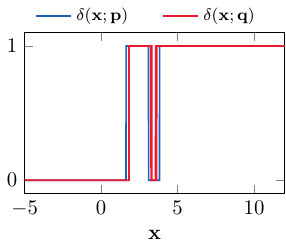}}
 \hfill
 \subfloat[\label{fig:postvar_np}Posterior variance under the nominal (\textcolor{my_blue}{{$\pmb -$}}) and least favorable model (\textcolor{my_red}{{$\pmb -$}}).]{\includegraphics{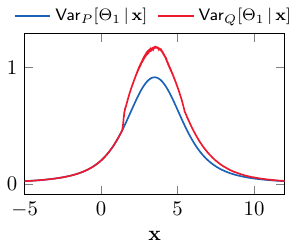}}
 \caption{\label{fig:res_np}Least-favorable densities and resulting decision rules under the NP-like formulation.}
\end{figure}
\subsection{Neyman-Pearson-like Formulation}\label{subsec:num_NP}
In this section, the \gls{NP}-like formulation is inspected.
More precisely, we set $\detErrMax[0]=0.05$ and $\detErrMax[0]=0.3$.
Additionally, we chose $\underbar{c}=0.8$ and $\bar{c}=1.5$ for the neighborhood.
For both, the optimal and the minimax optimal solution, we set $\estCost[1]=1$, which yields $\detCost[0]=0.522$, $\detCost[1]=0.901$ for the optimal and $\detCost[0]=0.613$, $\detCost[1]=1.231$ for the minimax optimal procedure.

The \glspl{lfd} as well as their derived quantities like the decision rule or the posterior variance are shown in \cref{fig:res_np}.
In \cref{fig:lfd_det_np}, the  \glspl{lfd} used for calculating the error probabilities are shown.
As expected, the \gls{lfd} under $\Hyp[0]$ shifts a lot of probability mass to the right, i.e., towards $\Hyp[1]$.
Similarly, the \glspl{lfd} under $\Hyp[1]$ shift most probability mass to the left, i.e. towards $\Hyp[0]$.
However, this effect is counteracted by the fact that shifting probability mass towards $\obs\approx3.5$ leads to a very high posterior variance, and, hence, to a high cost of accepting $\Hyp[1]$.
From \cref{fig:lfd_est_np}, one can see that the densities for $\param[]=1$ and $\param[]=6$ shift most of their mass towards $\obs\approx3.5$, where the posterior variance has its maximum.
Contrary to this, the density for $\param[]=3.5$ tries to maximize its variance within the band.
It can further be seen that the decision rules in \cref{fig:dec_np} have almost the same shape.
For the posterior variances as shown in \cref{fig:postvar_np}, it can be observed that the posterior variance of the nominal and the minimax procedure coincide for small values of $\obs$.
For larger values of $\obs$, the posterior variance is larger than the one of the optimal procedure, which can be explained as follows.
For  small values of $\obs$, the null hypothesis is accepted and, hence, in this region a maximization of the posterior variance would not lead to an increase of the overall cost.

In \cref{tbl:np}, the results of the Monte Carlo simulations for both, the optimal and the minimax optimal procedure are summarized for different scenarios.
More precisely, both procedures are evaluated using the nominal distributions, the \glspl{lfd} $\boldsymbol q_{\text{D}}$, the \gls{lfd} that maximizes the Type-I error probability together with the \glspl{lfd} that maximize the \gls{mse}.
First, it can be observed that the optimal and the minimax optimal procedure hit the targeted error probabilities exactly if the simulation is performed under $\allDens$ and $\boldsymbol q_{\text{D}}$, respectively.
Next, the empirical error probabilities of the optimal procedure with $\boldsymbol q_{\text{D}}$ severely exceed the targeted error probabilities, which underpins the need for minimax optimal procedures.
It can further be seen that the \gls{mse} increases when other distributions than the nominal ones are used.
Interestingly, the minimax optimal procedure always yields a larger \gls{mse} than the optimal one.
This might look counterintuitive as the minimax optimal procedure should \emph{minimize the maximum \gls{mse}}.
However, the idea of the \gls{jde} formulation as in \cref{eq:non_rob_prob_NP} is to slightly increase the Type-II error probability to leave room for further reducing the estimation error level.
As both error probabilities are maximized for the minimax procedure, this leaves less room for minimizing the \gls{mse} than it is the case for the optimal procedure.

\begin{table}[tp!]
    \centering
    \caption{\gls{NP}-like formulation: results of Monte Carlo simulation.}
    \label{tbl:np}
    \begin{tabular}{l|c|c|c|c}
        & \multicolumn{2}{c|}{error probabilities} &\\
        & type-I & type-II & \gls{mse} & $\NPcost$\\
        \toprule
        opt. with $\allDens$ & $0.050$ & $0.305$ &  $0.455$ & $0.378$ \\
        opt. with $\boldsymbol q_{\text{D}}$ & $0.074$ & $0.357$ & $0.477$ &  $0.419$ \\
        opt. with $\boldsymbol q_{\text{D},0}/\boldsymbol q_{\text{E}}$  & $0.074$ & $0.291$ & $0.617$ & $0.459$\\
        \midrule
        minimax with $\allDens$ & $0.033$  & $0.257$  & $0.511$  & $0.422$  \\
        minimax with $\boldsymbol q_{\text{D}}$ & $0.050$  & $0.305$  & $0.540$  & $0.470$\\
        minimax with $\boldsymbol q_{\text{D},0}/\boldsymbol q_{\text{E}}$ & $0.050$ & $0.233$ & $0.690$ & $0.501$ 
    \end{tabular}        
\end{table}

      \section{Conclusions}\label{sec:conclusion}
The problem of joint detection and estimation under distributional uncertainties has been investigated under a Bayesian and a \gls{NP}-like formulation.
We have shown that the optimal policy induces an $f$-similarity as a statistical similarity measure that is to be maximized by the worst case distributions; irrespective of the type of uncertainty sets.
To design minimax optimal procedures under the band model, we have adopted existing algorithms to increase convergence speed while maintaining numerical stability.
To supplement the proposed theory, numerical results have been provided for the Bayesian and for the \gls{NP}-like formulation. 
 \appendix
  \subsection{Proof of Theorem \ref{cor:BayesCostScalarArgProp}}\label{proof:propObjBayes}
First, it is shown that $\breve D_i^\text{B}(\bar{s}_{0,\param[0]},\bar{s}_{1,\param[1]})$, $i\in\{0,1\}$, are continuous and homogeneous in the arguments $\bar{s}_{0,\param[0]}$ and $\bar{s}_{1,\param[1]}$.
From the definition of $\breve D_i^\text{B}(s_0,s_1)$, one can see that it is a continuous function of $\bar{s}_{0,\param[0]}$ and $\bar{s}_{1,\param[1]}$. It holds that
\begin{align*}
    \breve D_i^\text{B}(\arbConst \bar{s}_{0,\param[0]}, \arbConst\bar{s}_{1,\param[1]}) = \arbConst \breve D_i^\text{B}(\bar{s}_{0,\param[0]}, \bar{s}_{1,\param[1]})
\end{align*}
for $\arbConst>0$.
Furthermore, we can state that 
\begin{align*}
    \lim_{\arbConst\rightarrow0}\breve D_i^\text{B}(\arbConst \bar{s}_{0,\param[0]}, \arbConst\bar{s}_{1,\param[1]}) = \lim_{\arbConst\rightarrow0}\arbConst\breve D_i^\text{B}(\bar{s}_{0,\param[0]}, \bar{s}_{1,\param[1]}) = 0\,.
\end{align*}
Since the minimum operator preserves these properties, they hold for $\rho^\text{B}(\cdot)$ as well.
As stated in the beginning of \cref{sec:Bayes}, $\BayesCost(\policy, p_{0, \param[0]}, p_{1, \param[1]})$ is concave in the densities, and, hence the function $\rho^\text{B}(\cdot)$ is also concave in its arguments. \hfill\IEEEQED

\subsection{Proof of Theorem \ref{theo:minimax_NP}}\label{proof:minimax_NP}
First, for a given policy $\policyOpt$, we can see that
    \begin{align*}
      &\phantom{{}={}}\max_{\substack{\boldsymbol P_0^\text{D} \in \mathbb{P}_0\,,\;\boldsymbol P_0^\text{E} \in \mathbb{P}_0 \\ \boldsymbol P_1^\text{D} \in \mathbb{P}_1\,,\;\boldsymbol P_1^\text{E} \in \mathbb{P}_1}}
        \;D_{\rho^\text{NP}}(\boldsymbol P_0^\text{D}, \boldsymbol P_1^\text{D}, \boldsymbol P_0^\text{E}, \boldsymbol P_1^\text{E}) \\
        & = \max_{\substack{\boldsymbol P_0^\text{D} \in \mathbb{P}_0\,,\;\boldsymbol P_0^\text{E} \in \mathbb{P}_0 \\ \boldsymbol P_1^\text{D} \in \mathbb{P}_1\,,\;\boldsymbol P_1^\text{E} \in \mathbb{P}_1}}\!\! \left(
        \sum_{i=0}^1 p(\Hyp)\biggl(\detCost[i]\detErr[i](\policy,\boldsymbol p_i^\text{D}) + \estCost[i]\estErr[i](\policy,\boldsymbol p_i^\text{E})\biggr)
        \right)
    \end{align*}    
    implies that
    \begin{align*}
        \max_{\boldsymbol p_i^\text{D} \in \mathbb{P}_i} \;\detErr(\policy, \boldsymbol p^\text{D}_i)\,,\quad
        \max_{\boldsymbol p_i^\text{E} \in \mathbb{P}_i} \;\estErr(\policy, \boldsymbol p_i^\text{E})\,,\quad i\in\{0,1\}
    \end{align*}
    as \emph{for a given policy}, there is no coupling in the densities of the individual performance measures.

    Assume that there exists a policy $\policy^\circ = (\dec^\circ, \est[0]^\circ, \est[1]^\circ)$ such that
    \begin{align*}
        \NPcost(\policy^\circ, \boldsymbol q_0^\text{D}, \boldsymbol q_1^\text{D}, \boldsymbol q_0^\text{E}, \boldsymbol q_1^\text{E}) < \NPcost(\policyOpt, \boldsymbol q_0^\text{D}, \boldsymbol q_1^\text{D}, \boldsymbol q_0^\text{E}, \boldsymbol q_1^\text{E})
    \end{align*}
    and
    \begin{align*}
        \detErr(\policy^\circ, \detDensSetLFD) \leq   \detErrMax\,, \quad i\in\{0,1\}.
    \end{align*}
    From the above, we can conclude that
    \begin{align*}
        &\phantom{{}={}} \NPcost(\policyOpt, \allDensLFD) + \sum_{i=0}^1 \detCost p(\Hyp) \detErr(\policyOpt, \detDensSetLFD) \\
        & = \NPcost(\policyOpt, \allDensLFD) + \sum_{i=0}^1 \detCost p(\Hyp) \detErrMax \\
        & \geq \NPcost(\policy^\circ, \allDensLFD) + \sum_{i=0}^1 \detCost p(\Hyp) \detErr(\policy^\circ, \detDensSetLFD)
    \end{align*}
    which contradicts the assumption that the policy stated in \cref{theo:optPolicyNP} solves \cref{eq:NP_obj_unconstr}.
    It is left to show that \cref{eq:detErr_NP} could be obtained by solving \cref{eq:optCostCoeff}.
    First, it can be shown that
    \begin{align*}
        \frac{\partial}{\partial \detCost[j]} D_i^\text{NP} = \begin{cases}
            p(\Hyp[1-i])\detDens[1-i](\obs\given\Hyp[1-i]) & j=1-i\,, \\
            0 & \text{else}\,.
        \end{cases}
    \end{align*}
    With
    \begin{align*}
        \NPcostUnconstr(\obs) = (1-\dec(\obs))D_0^\text{NP}(\obs) + \dec(\obs)D_1^\text{NP}(\obs)
    \end{align*}
    we can, under mild assumptions, conclude that
    \begin{align*}
        \frac{\partial}{\partial \detCost[0]} \NPcostUnconstr(\obs) = \int \dec(\obs) p(\Hyp[0])p(\obs\given\Hyp[0])\dInt\obs = p(\Hyp[0])\detErr[0]\,.
    \end{align*}
    Then, it follows that the objective in \cref{eq:optCostCoeff} attains an extremum at $\detErr[0] = \detErrMax[0]$ if $\detCost[0]$ > 0.
    If $\detCost[0]=0$, it acts as a slack variable and it follows that $\detErr[0]<\detErrMax[0]$.
With the same line of arguments, it can be shown that a similar relationship holds for $\detCost[1]$.
    \hfill\IEEEQED
   \subsection{Gradient of the Objective}\label{sec:gradient}
In this section, we use the short-hand notations $\boldsymbol s=(\boldsymbol s_{0,\param[0]}, \boldsymbol s_{1,\param[1]})$ or $\boldsymbol s=(\boldsymbol s_{0,\param[0]}^\text{D}, \boldsymbol s_{1,\param[1]}^\text{D}, \boldsymbol s_{0,\param[0]}^\text{E}, \boldsymbol s_{1,\param[1]}^\text{E})$ to keep the notation compact.
First, the Bayesian formulation as used in \cref{sec:Bayes} is considered.
The derivative of the cost for accepting hypothesis $\Hyp$ for $i\in\{0,1\}$ is given by
\begin{align*}
    \frac{\partial}{\partial \boldsymbol s_{i,\param}} \tilde D_i^\text{B}(\boldsymbol s)  
    & = \mu_i p(\Hyp) \boldsymbol a_i
    - 2\mu_i p(\Hyp) \frac{\boldsymbol b_i^\top s_{i,\param}}{\boldsymbol c_i^\top\boldsymbol s_{i,\param}}\boldsymbol b_i \\
     & \quad + \mu_i p(\Hyp)\frac{\bigl(\boldsymbol b_i^\top\boldsymbol s_{i,\param}\bigr)^2}{\bigl(\boldsymbol c_i^\top\boldsymbol s_{i,\param}\bigr)^2} \boldsymbol c_i \\
\frac{\partial}{\partial \boldsymbol s_{1-i,\param[1-i]}} \tilde D_i^\text{B}(\boldsymbol s) & = \lambda_{1-i}p(\Hyp[1-i])\boldsymbol c_{1-i}
\end{align*}

From the definition of the objective
\begin{align*}
    f(\boldsymbol s) = -\min\{\tilde D_0^\text{B}(\boldsymbol s), \tilde D_1^\text{B}(\boldsymbol s)\}\,,
\end{align*}
we can directly conclude that the final gradient becomes
\begin{align*}
\frac{\partial f(\boldsymbol s)}{\partial \boldsymbol s_{i,\param}} = -\begin{cases}
     \frac{\partial}{\partial \boldsymbol s_{i,\param}} \tilde D_i^\text{B}(\boldsymbol s) &\tilde D_i^\text{B}(\boldsymbol s) < \tilde D_{1-i}^\text{B}(\boldsymbol s) \,, \\
     \frac{\partial}{\partial \boldsymbol s_{i,\param}} \tilde D_{1-i}^\text{B}(\boldsymbol s) & \tilde D_i^\text{B}(\boldsymbol s) > \tilde D_{1-i}^\text{B}(\boldsymbol s)\,.
    \end{cases}
\end{align*}

Second, when using the \gls{NP}-like formulation, the cost for accepting hypothesis $\Hyp$ is given by
\begin{align*}
    \frac{\partial}{\partial \boldsymbol s_{i,\param}^\text{E}} \tilde D_i^\text{NP}(\boldsymbol s) & 
    = \mu_i p(\Hyp) \boldsymbol a_i
    - 2\mu_i p(\Hyp) \frac{\boldsymbol b_i^\top s_{i,\param}^\text{E}}{\boldsymbol c_i^\top\boldsymbol s_{i,\param}^\text{E}}\boldsymbol b_i\\
     & \phantom{{}={}} + \mu_i p(\Hyp)\frac{\bigl(\boldsymbol b_i^\top\boldsymbol s_{i,\param}^\text{E}\bigr)^2}{\bigl(\boldsymbol c_i^\top\boldsymbol s_{i,\param}^\text{E}\bigr)^2} \boldsymbol c_i \\
    \frac{\partial}{\partial \boldsymbol s_{1-i}^\text{D}} \tilde D_i^\text{NP}(\boldsymbol s) & 
    = \lambda_{1-i}p(\Hyp[1-i])\boldsymbol c_{1-i} \\
    \frac{\partial}{\partial \boldsymbol s_{i,\param}^\text{D}} \tilde D_i^\text{NP}(\boldsymbol s) & = 0  \\   
    \frac{\partial}{\partial \boldsymbol s_{1-i,\param[1-i]}^\text{E}} \tilde D_i^\text{NP}(\boldsymbol s) & = 0\\
\end{align*} This results in an overall gradient of the objective:
\begin{align*}
    \frac{\partial f(\boldsymbol s)}{\partial \boldsymbol s_{i,\param}^\text{E}}  = -\begin{cases}
     \frac{\partial}{\partial \boldsymbol s_{i,\param}^\text{E}} \tilde D_i^\text{NP}(\boldsymbol s) & \tilde D_i^\text{NP}(\boldsymbol s) < \tilde D_{1-i}^\text{NP}(\boldsymbol s)\,, \\
     0 &\tilde D_i^\text{NP}(\boldsymbol s) > \tilde D_{1-i}^\text{NP}(\boldsymbol s)\,.
    \end{cases}
\end{align*}
\begin{align*}
    \frac{\partial f(\boldsymbol s)}{\partial \boldsymbol s_{i,\param}^\text{D}}  = -\begin{cases}
     \frac{\partial}{\partial \boldsymbol s_{i,\param}^\text{D}} \tilde D_{1-i}^\text{NP}(\boldsymbol s) &\tilde D_i^\text{NP}(\boldsymbol s) > \tilde D_{1-i}^\text{NP}(\boldsymbol s)\,, \\
     0 &\tilde D_i^\text{NP}(\boldsymbol s) < \tilde D_{1-i}^\text{NP}(\boldsymbol s)\,.
    \end{cases}
\end{align*}

However, as the gradient causes practical implementation issues when finding the inverse of the gradient, we replaced the conventional minimum operator by a \emph{softmin} operator of the form
\begin{align*}
    \softmin\{a, b\} := \xi\cdot\Biggl( \log\Bigl(\exp\bigl(\xi a^{-1}\bigr) + \exp\bigl(\xi b^{-1}\bigr)\Bigr)\Biggr)^{-1}\,,
\end{align*}
where $\xi$ is a scale variable that controls the steepness of the function.

\bibliographystyle{IEEEtran}
\bibliography{references} 

\end{document}